\documentclass[12pt]{article}

\usepackage[dvips]{graphicx}
\usepackage{epsfig}
\usepackage{amsmath,amsfonts,amssymb,amsthm}
\usepackage{verbatim}
\usepackage{psfrag}
\usepackage{bm}
\usepackage{bbm}
\usepackage[square,comma,sort&compress,numbers]{natbib}
\usepackage{color}
\usepackage{slashed}

\usepackage{epsf,epsfig}
\usepackage{graphics}

\begin{document}
\thispagestyle{empty}

\begin{flushright}
{
\small
TTK-12-02
}
\end{flushright}

\vspace{0.4cm}
\begin{center}
\Large\bf\boldmath
Leptogenesis from Additional Higgs Doublets
\unboldmath
\end{center}

\vspace{0.4cm}

\begin{center}
{Bj\"orn~Garbrecht}\\
\vskip0.2cm
{\it Institut f\"ur Theoretische Teilchenphysik und Kosmologie,\\ 
RWTH Aachen University, 52056 Aachen, Germany}\\
\vskip1.4cm
\end{center}

\begin{abstract}
Leptogenesis may be induced by the mixing of extra Higgs doublets with experimentally
accessible masses. This mechanism relies on diagrammatic cuts that are kinematically
forbidden in the vacuum but contribute at finite temperature.
A resonant enhancement of the asymmetry occurs generically
provided the dimensionless Yukawa and self-interactions are suppressed compared to
those of the Standard Model Higgs field. This is in contrast to typical scenarios of
Resonant Leptogenesis, where the asymmetry is enhanced by imposing a degeneracy of
singlet neutrino masses.
\end{abstract}



\section{Introduction}

Experimentally accessible $CP$-violating effects, such as the mixing of
neutral mesons, are often resonantly enhanced. This suggests that the
matter-antimatter asymmetry of the Universe may have emerged from
resonant mixing as well. The enhancement is maximized
when the mixing states are nearly degenerate in their masses
and couplings.
For example, Resonant Leptogenesis~\cite{Covi:1996wh,Flanz:1996fb,Pilaftsis:1997dr,Pilaftsis:1997jf,Pilaftsis:2003gt}
relies on almost
mass-degenerate singlet neutrinos $N$, where the masses are
of Majorana type and violate lepton number.
The degeneracy may result from an approximate global symmetry or from parametric tuning. One reason for introducing it is to allow for low
reheat temperatures. Barring resonant enhancement, the observed
asymmetry can only arise at temperatures above $10^9\,{\rm GeV}$,
which lead to the overproduction of unwanted thermal relics in some
Particle Physics scenarios.

In the most generic realizations, the couplings
of the singlet neutrinos are so small that they escape detection in
colliders, even if their masses are of
Electroweak scale or within an experimentally accessible range above.
(See, however, Ref.~\cite{Pilaftsis:2005rv}.)
It is therefore interesting to consider the question whether
particles that are not gauge singlets can lead to a resonantly
enhanced asymmetry as well. Such particles could be found in collider
experiments.

Obvious candidates are additional Higgs doublets $\phi$.
These fields combine to gauge singlets
with the Standard Model (SM) lepton doublets $\ell$, such that they
can couple to a singlet neutrino at the renormalizable level.
Due to its arbitrarily
small couplings, the singlet may deviate
from equilibrium in the Early Universe, thus breaking $T$ invariance,
which is necessary in order to make the $CP$-violation needed for
Leptogenesis effective, what would be barred by the $CPT$-theorem
otherwise. The masses and couplings of the Higgs doublets do
not violate lepton number, but generically violate
lepton flavor. This implies that before washout,
no lepton asymmetry, but only a flavor asymmetry can be generated.
The flavor asymmetries suffer different amounts of washout, such that
eventually, a lepton asymmetry emerges.

With the basic mechanism outlined, its quantitative description
should address the following key questions:
\begin{itemize}
\item
Concerning $CP$-violation from
mixing, the Higgs doublets assume the role of the singlet neutrinos
in conventional Leptogenesis. Gauge interactions keep
these doublets close to thermal equilibrium. The deviation
from equilibrium is mediated by loops involving the singlet neutrinos.
Does this induce a $CP$-violating bias in the production of leptons? To answer
this, an analysis with accurate account for real intermediate
states must be performed. For this purpose,
we use the Closed-Time-Path (CTP)
formalism. The lepton production rate follows from the imaginary part of
the diagram in Figure~\ref{fig:SigellCPV}, and
the correct counting of real and virtual states is
incorporated automatically~\cite{Buchmuller:2000nd,De Simone:2007rw,Garny:2009rv,Garny:2009qn,Anisimov:2010aq,Garny:2010nj,Beneke:2010wd,Beneke:2010dz,Garny:2010nz,Garbrecht:2010sz,Anisimov:2010dk}.
\item
When the masses of the extra doublets $\phi$ are of Electroweak or
${\rm TeV}$ scale and
$N$ is much heavier, the
relevant cut that corresponds to the process $\phi\to \bar\ell+N$
is kinematically forbidden in the
vacuum. In a thermal background however, it receives contributions from
the distribution functions of $N$ and $\ell$. Again, this
can be reliably calculated using CTP methods.
\item
The resonance for almost degenerate Higgs-doublets should be limited by their
widths. We show that the relevant contributions arise from Higgs-flavor violating
Yukawa and scalar self-interactions, while the universal flavor conserving gauge
interactions do not limit the resonant enhancement. Note that a partial cancellation
of the widths that limit the enhancement is familiar from
conventional Resonant Leptogenesis~\cite{Buchmuller:1997yu,Anisimov:2005hr,Garbrecht:2011aw,Garny:2011hg}.
\end{itemize}

\section{The Model and the Mechanism}

The present proposal is realized by
the Lagrangian
\begin{align}
{\cal L}=&\frac12 \bar \psi_N({\rm i}\slashed\partial - M_N)\psi_N
+\bar \psi_\ell{\rm i}\slashed\partial\psi_\ell
+\sum\limits_k(\partial_\mu \phi_k)(\partial^\mu \phi_k)
\notag\\
-&
\sum\limits_{kl} M^2_{\phi kl}\phi_k^*\phi_l
-\sum\limits_{mk}(Y_{mk}\bar \psi_N \phi_k P_{\rm L}\psi_{\ell m}+{\rm h.c.})
\,.
\end{align}
The spinor $\psi_ N$ represents the singlet Majorana neutrino $N$,
$\psi_{\ell m}$ with $m=1,2$ two flavors of SM lepton
doublets $\ell$
and $\phi_k$ with $k=1,2$ are two extra Higgs doublets,
besides the one of the SM. The matrix $M^2_\phi$ is
diagonal, what can be achieved by field redefinitions, and indices
of the ${\rm SU}(2)$ gauge group are suppressed. We take $M^2_\phi$
as an effective mass, that readily includes thermal corrections.

\begin{figure}[t!]
\begin{center}
\epsfig{file=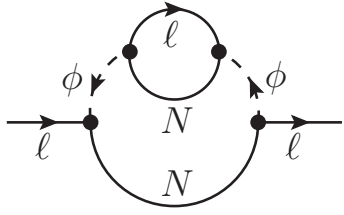,width=5.5cm}
\end{center}
\vskip-.5cm
\caption{\label{fig:SigellCPV}
The imaginary part of this Feynman diagram
yields the $CP$-violating production rate
of lepton flavors.}
\end{figure}

Using the CTP approach, Leptogenesis can be described by the
simple network of equations
\begin{align}
\label{kineq:netw}
\frac{dq_{\ell i}}{d\eta}&=S_{\ell i}+W_{\ell_i}q_{\ell i}\,,
\qquad
\frac{d f_{N}(\mathbf k)}{d\eta}={\cal C}_N(\mathbf k)\,.
\end{align}
Here, $q_{\ell i}$ is the comoving charge density of leptons of
flavor $i$ and $f_X$ is the comoving distribution
function for the particle $X$, 
$f_X^{\rm eq}$ are equilibrium distributions
and $\eta$ is
the conformal time. When $T\gg M_{\phi}$,
where $T$ is the temperature,
the collision term for singlet neutrinos is~\cite{Beneke:2010wd}
\begin{align}
\label{C:N}
{\cal C}_N(\mathbf k)
=&\frac{\sum_{mk}|Y_{mk}|^2}{\sqrt{\mathbf k^2+M_N^2}}
\int\frac{d^3 p\,d^3 q}{(2\pi)^2 4|\mathbf p|\,|\mathbf q|}\delta^4(k-p-q)
2k\cdot p
\notag\\
\times
&
[1-f_{\ell_m}^{\rm eq}(\mathbf p)+f_{\phi_{kk}}^{\rm eq}(\mathbf q)]\times[f^{\rm eq}_N(\mathbf k)-f_N(\mathbf k)]\,,
\end{align}
and the washout term for the leptons~\cite{Beneke:2010wd}
\begin{align}
\label{washout}
W_{\ell i}=&-\sum\limits_k |Y_{ik}|^2
\int\frac{d^3 k\,d^3p\,d^3 q\,2p\cdot k}{(2\pi)^5
8 |\mathbf k|\sqrt{\mathbf p^2+M_N^2}|\mathbf q|}
\delta^4(p-k-q)
[f_{\phi_{kk}}^{\rm eq}(\mathbf q)+f_N(\mathbf p)]
\frac{12 T^{-3}{\rm e}^{|\mathbf k|/T} }{({\rm e}^{|\mathbf k|/T}+1)^2}\,.
\end{align}
Within the phase-space integrals, we have approximated
here $\ell$ and $\phi$ as massless, such that we may
substitute
$f_{\ell_1}^{\rm eq}=f_{\ell_2}^{\rm eq}\equiv f_{\ell}^{\rm eq}$
and $f_{\phi_{11}}^{\rm eq}=f_{\phi_{22}}^{\rm eq}\equiv f_{\phi}^{\rm eq}$
by the Fermi-Dirac and Bose-Einstein distributions for massless particles.
Of special interest is, of course, the $CP$-violating contribution
to the lepton collision term
\begin{align}
\label{coll:l}
{\cal C}&^{\rm CPV}_{\ell i}(\mathbf k)=
\frac{{\rm Im}[Y_{i1}Y^*_{j1}Y_{j2}Y^*_{i2}]}{M_{\phi11}^2-M_{\phi22}^2}
\int\frac{dk^0d^4p\,d^4q}{(2\pi)^9}
\notag\\
\times&[{\rm i}S_N^>(p+k){\rm i}S_{\ell i}^<(k)
-
<\leftrightarrow>
]
[{\rm i}\Delta_{\phi_{11}}^<(p)
P_{\rm L}S_{\ell j}^<(-p-q)
{\rm i}S_N^<(q)
-
<\leftrightarrow>
]+1\leftrightarrow2\,,
\end{align}
where $i\not=j$.
This term can be derived using the methods of Ref.~\cite{Beneke:2010wd},
and it
enters in Eqs.~(\ref{kineq:netw}) as
$S_{\ell i}=\int d^3k\,{\cal C}^{\rm CPV}_{\ell i}/(2\pi)^3$.
Note that ${\cal C}_{\ell 1}^{\rm CPV}=-{\cal C}_{\ell 2}^{\rm CPV}$,
as a consequence of lepton number conservation in the diagram of
Figure~\ref{fig:SigellCPV}. Eventually, the total asymmetry arises from
different washout of the two flavors~\cite{Endoh:2003mz,Pilaftsis:2005rv,Abada:2006fw,Nardi:2006fx}.
The fermionic and scalar Wightman functions $S^{<,>}_X$
and $\Delta^{<,>}_\phi$ describe the phase-space distributions
of the various quasi-particles and are given {\it e.g.}
in Ref.~\cite{Beneke:2010wd}. For the
present discussion, we note that in thermal equilibrium, the
Kubo-Martin-Schwinger (KMS) relations $S_X^>(p)=-{\rm e}^{p^0/T}S_X^<(p)$
and $\Delta_\phi^>(p)={\rm e}^{p^0/T}\Delta_\phi^<(p)$ hold,
which imply that the
last factor in Eq.~(\ref{coll:l}) vanishes in equilibrium. However,
when $N$
deviates from equilibrium, this term is proportional to
$\delta f_N=f_N-f_N^{\rm eq}$.

The $CP$-violating source, Eq.~(\ref{coll:l}),
corresponds to the insertion of a
single loop in the Higgs propagator in Figure~\ref{fig:SigellCPV}. The question of
how finite width limits the resonance can be answered when resumming
these insertions as well as those from the additional
interactions. When we denote the resummed propagator
of the mixing Higgs fields by $D_\phi$, the source term becomes
\begin{align}
\label{source:resummed}
S_{\ell i}=&
\sum\limits_{^{k,l}_{k\not=l}}
Y^*_{ik}Y_{il}
\int
\frac{d^4k\,d^4p\,d^4q\,\delta^4(k+p-q)}{(2\pi)^8}
{\rm tr}
\left[
P_{\rm R}{\rm i}S_N^>(q){\rm i}S^<_{\ell i}(k)
-<\leftrightarrow>
\right]
{\rm i}D_{\phi kl}(p)\,,
\end{align}
where $i\not=j$.
We drop the superscripts $<,>$ on $D$, since we consider here only its
off-diagonal components, for which $D^<\equiv D^>$ at the leading,
non-vanishing order. Moreover,
${\rm i}D_{\phi_{21}}=({\rm i}D_{\phi_{12}})^*$.
The off-diagonal components
of the
resummed propagator can be obtained from the kinetic equations~\cite{Garbrecht:2011aw,Garny:2011hg},
\begin{align}
\label{kin:eq}
&2k^0\partial_\eta {\rm i}D_{\phi 12}
+{\rm i}(M^2_{\phi 11}-M^2_{\phi 22}){\rm i}D_{\phi 12}
\notag\\
=&-\frac12 {\rm i}(\Pi^{\slashed{\rm fl}>}_{\phi12}+\Pi^{Y>}_{\phi12}+\Pi^{g>}_{\phi12})
{\rm i}(\Delta^<_{\phi_{11}}+\Delta^<_{\phi_{22}})
-
\frac12 \sum\limits_k {\rm i}(\Pi^{\slashed{\rm fl}>}_{\phi_{kk}}+\Pi^{Y>}_{\phi_{kk}}
+\Pi^{g>}_{\phi_{kk}}){\rm i}D_{\phi 12}
-
<\leftrightarrow>\,,
\end{align}
as a perturbation to the diagonal Wightman functions
$\Delta^{<,>}_{\phi_{ii}}$, that are of equilibrium form.
The Higgs-flavor violating contributions to the self-energy, that are mediated
by additional Yukawa couplings $y$ and scalar self-interactions
$\lambda$ are
given by $\Pi_\phi^{\slashed{\rm fl}<,>}$, the contributions
mediated by the couplings $Y$ by $\Pi_\phi^{Y<,>}$ and those
from gauge couplings $g$ by $\Pi_\phi^{g<,>}$. The latter are flavor
conserving in the sense that when integrating Eq.~(\ref{kin:eq})
over $d^4k$, the terms
involving $\Pi_\phi^{g<,>}$ vanish. This is shown in Ref.~\cite{Beneke:2010dz} for
flavor-coherent fermions, but the argument can be directly applied
to the present scalar case. Unfortunately, this argument does not solve
Eq.~(\ref{kin:eq}) in general. However, when the lifetime of the virtual
Higgs $1/|M_{\phi{11}}-M_{\phi{22}}|$ is much larger than $gT$, the
time-scale of kinetic equilibration, the off-diagonal components of
$D$ follow a kinetic equilibrium distribution~\cite{Beneke:2010dz},
\begin{align}
\label{D:resummed}
{\rm i}D_{\phi 12}(p)= 2\pi \delta(p^2-\bar M^2_\phi)\frac{\mu_{\phi{12}}}{T}
\frac{{\rm sign}(p^0){\rm e}^{|p^0|/T}}{({\rm e}^{|p^0|/T}-1)^2}\,,
\end{align}
where $\bar M_\phi=|M_{\phi{11}}-M_{\phi{22}}|/2$ and for $M_\phi\ll T$,
\begin{align}
\label{q12:fromsource}
q_{\phi 12}=\frac{\mu_{\phi 12} T^2}3=2\int\frac{d^4 k}{(2\pi)^4}k^0{\rm i}D_{\phi_{12}}(k)\,.
\end{align}
Note that in kinetic equilibrium, the terms in Eq.~(\ref{kin:eq})
that involve $\Pi_\phi^{g<,>}$ readily cancel before integration over
$d^4 k$.
From Eq.~(\ref{kin:eq}), when neglecting the derivative
with respect to $\eta$, it then follows
\begin{align}
\label{q12:explicit}
q_{\phi 12}\!=&
\frac
{4{\rm i}\sum_jY_{j1}Y_{j2}^*}
{M^2_{\phi 11}-M^2_{\phi 22}+{\rm i} T \bar \Gamma^{\slashed{\rm fl}}}
\!\int\frac{d^3 k\,d^3p\,d^3q\,\delta^4(\!k+\!p\!-\!q)}{(2\pi)^5 8 |k|\,|p|\,\sqrt{|\mathbf q|^2+M_N^2}}
\notag\\
\times&
k^0
2p\cdot q[1-f^{\rm eq}_\ell(\mathbf p)+f^{\rm eq}_\phi(\mathbf k)]\delta f_N(\mathbf q)\,,
\end{align}
where
$\bar\Gamma^{\slashed{\rm fl}}$
is a weighted average of $\Pi_\phi^{\slashed{\rm fl}<,>}$ and
$\Pi_\phi^{Y<,>}$, which we estimate below.
This limits the resonant enhancement of the asymmetry.
The important point is that the flavor-conserving gauge interactions do
not contribute to $\bar\Gamma^{\slashed{\rm fl}}$ at leading order.

\section{Lepton Asymmetry}

Though to this end, the discussion has been more general, we
now restrict to a phenomenologically interesting region of parameter
space where analytic approximations of good accuracy are available.
In the strong washout regime, the right handed neutrino is non-relativistic,
{\it i.e.} $M_N\gg T$, when the asymmetry freezes out.
Moreover, we assume that $M_{\phi ii}\ll M_N$. Then, we can
neglect the masses of the Higgs particles in the phase space integrals and
only keep them in the resonant enhancement factors. As $M_N\gg T$,
the decay products are of energy much larger than $T$ as well. Therefore,
we may replace the quantum-statistical
distribution functions by Maxwell distributions.


Moreover, we describe the deviation of the right-handed neutrino from equilibrium
by a pseudo-chemical potential,
$\delta f_N(\mathbf k)\approx\exp(\sqrt{\mathbf k^2+M_N^2})\mu_N/T$. Even though
there are no interactions that drive $N$ toward kinetic equilibrium,
this empirically proves to be a good approximation~\cite{Beneke:2010wd,Basboll:2006yx}.
We then obtain
the off-diagonal charges in the mixing Higgs-system from Eq.~(\ref{q12:explicit}),
\begin{align}
q_{\phi 12}=\frac{{\rm i} \sum_j Y_{j1}Y_{j2}^*}{M_{\phi 11}^2-M_{\phi 22}^2}
\frac{\mu_N M_N^{7/2}T^{1/2}}{16\sqrt2\pi^{5/2}}
{\rm e}^{-M_N/T}\,,
\end{align}
where we
approximate $\bar\Gamma^{\slashed{\rm fl}}\ll|M_{\phi11}-M_{\phi22}|$,
which turns out as generic.
Substitution into Eqs.~(\ref{D:resummed},\ref{q12:fromsource}) and~(\ref{source:resummed}) yields
\begin{align}
\label{S:Maxwell}
S_{\ell i}
=\frac{3 \mu_N M_N^6}{512\pi^5 T}\frac{{\rm Im}[Y_{i1}Y^*_{j1}Y_{j2}Y^*_{i2}]}{M_{\phi{11}}^2-M_{\phi{22}}^2}
{\rm e}^{-2M_N/T}\,.
\end{align}

Now we are set to follow the standard routine for calculating the
asymmetry in the strong washout regime~\cite{Kolb:1983ni,Buchmuller:2004nz}. The expanding background is taken
account of following Ref.~\cite{Beneke:2010wd}, and
Eqs.~(\ref{kineq:netw})
simplify to
\begin{subequations}
\label{kineq:Y}
\begin{align}
\frac{dY_{\ell i}}{dz}&
=\bar S_{\ell i}(Y_N-Y_N^{\rm eq})+\bar W_{\ell i}Y_{\ell i}\,,
\\
\frac{dY_N}{dz}&=\bar {\cal C}_N(Y_N-Y_N^{\rm eq})\,,
\end{align}
\end{subequations}
where $Y_{\ell i}=4 q_{\ell i}/s$, $Y_N=2\int d^3k f_N(\mathbf k)/((2\pi)^3s)$ (the factors account for spin and $\rm{SU}(2)$ degrees of freedom),
$z=m_N/T$ and $s$ is the entropy density.
The integrated distributions and collision terms that occur here can be obtained
from Eqs.~(\ref{C:N},\ref{washout},\ref{S:Maxwell}) and are explicitly given by
\begin{subequations}
\begin{align}
Y_N^{\rm eq}&=2^{-1/2}\pi^{-3/2}a_{\rm R}^3 z^{3/2}{\rm e}^{-z}/s\,,
\\
\bar{\cal C}&=-\frac{1}{8\pi}{\rm tr}[YY^\dagger]\frac{a_{\rm R}}{M_N}z\,,
\\
\bar W_{\ell i}
&=
3\times 2^{-9/2} \pi^{-5/2}
\sum\limits_k
Y_{ik}Y^\dagger_{ki} \frac{a_{\rm R}}{M_N}z^{5/2}{\rm e}^{-z}
\,,
\\
\label{S:explicit}
\bar S_i&=\frac{3}{2^{13/2}\pi^{7/2}}
\frac
{{\rm Im}[Y_{i1}Y^*_{j1}Y_{j2}Y^*_{i2}]}
{M_{\phi{11}}^2-M_{\phi{22}}^2}
a_{\rm R}M_N z^{5/2}{\rm e}^{-z}\,,
\end{align}
\end{subequations}
where $a_{\rm R}=(1/2)m_{\rm Pl}(\pi^3g_\star/45)^{-1/2}$,
$m_{\rm Pl}=1.22\times10^{19}\,{\rm GeV}$ is the Planck mass
and
$g_\star$ the number of relativistic degrees of freedom.

The Eqs.~(\ref{kineq:Y}) can be formally integrated and then evaluated employing
Laplace's steepest descent method with the result
\begin{subequations}
\label{Yell:freezeout}
\begin{align}
\label{ass:fr}
Y_{\ell i}(z=\infty)&=
\frac{{\rm Im}[Y_{i1}Y^*_{j1}Y_{j2}Y^*_{i2}]}{{\rm tr[YY^\dagger]}}
\frac{135\,M_N^2}{64\pi^6 g_\star}\frac{1}{M_{\phi{11}}^2-M_{\phi{22}}^2}
\sqrt{z_{{\rm f}i}^{-1/2}{\rm e}^{z_{{\rm f}i}}}
{\rm e}^{-2z_{{\rm f}i}-\int\limits_{z_{{\rm f}i}}^\infty dz B_i z^{5/2}{\rm e}^{-z}}\,,
\,\\
B_i&=\sum\limits_k Y_{ik} Y^*_{ik}
\frac{9}{32\pi^4}\sqrt{\frac{5}{2g_\star}}\frac{m_{\rm Pl}}{M_N}\,,
\,\\
z_{{\rm f}i}&=-\frac{5}{2}W_{-1}\left((-{2}/{5})\times\left({2}/{B_i}\right)^{2/5}\right)\,,
\end{align}
\end{subequations}
where $W_{-1}$ is the lower branch of the product logarithm.
It should be clear that asymmetries $Y_{\ell i}$ of order $10^{-10}$ or larger
can easily be obtained, {\it cf.} Section~\ref{Sec:Obs:Bounds} for more details.
The temperature at freeze-out is given by $M_N/z_{{\rm f}i}$, which
is sufficiently accurately approximated when $B_i\gtrsim 1$, corresponding
to the strong washout regime.
In the exponent of Eq.~(\ref{ass:fr}), there is the term
$-2z_{{\rm f}i}$ rather than $-z_{{\rm f}i}$ for conventional
Leptogenesis~\cite{Kolb:1983ni,Buchmuller:2004nz},
because the cut that leads to the asymmetry
is kinematically forbidden in the vacuum, what gives an additional
Maxwell suppression. It is crucial that $B_1\not=B_2$,
such that a cancellation of the flavored asymmetries is avoided. (See however
Ref.~\cite{Chung:2008gv} for a loophole.)

Before applying the formula~(\ref{ass:fr}), we must consider
the mass difference and the flavor-violating
width, that limit the resonant enhancement. For
$M_\phi\ll T$, the relevant contributions
to the asymptotic (momenta much larger than $T$) Higgs mass are
\begin{align}
\label{M:decomp}
M^2_{\phi{ii}}=m^2_{\phi{ii}}+\left(\sum\limits_k \frac{\lambda_{ki} }{2}
+\sum\limits_j \frac{y_{ji}^2 }{12}
+\frac{3g_{\rm L}^2}{16}+\frac{g_Y^2}{16}\right)T^2\,,
\end{align}
where $\lambda_{ki}$ denotes the coupling to the Higgs doublet $k$,
$y_{ji}$ the coupling to the pair of chiral fermions $i$,
$m^2_{\phi{ii}}$ is the vacuum mass and
$g_{\rm L}$, $g_Y$ are the ${\rm SU}(2)$ and ${\rm U}(1)$ gauge couplings
of the SM.
Assuming that $g_{\rm L}^2,g_Y^2\gg Y^2,y^2,\lambda$, the leading contributions to the
flavor-violating width are
$\bar\Gamma^{\slashed{\rm fl}}={\cal O}({g_{\rm L}^2,g_Y^2}\times(y^2,Y^2,\lambda))\times T$.
This is because the damping of the flavor correlations of the
Higgs fields is kinematically
suppressed
by the small zero-temperature and thermal masses (compared to $T$) in
$1\leftrightarrow 2$ processes or by additional powers of couplings in
$2\leftrightarrow 2$ scattering processes. A full calculation
of $\Gamma^{\slashed{\rm fl}}$ is presently still very challenging, {\it cf.} the discussion
in Ref.~\cite{Anisimov:2010gy}.
In conclusion, the suppression of the resonance through thermal masses
is generically stronger than the effect of Higgs flavor-violating damping rates, when
barring accidental cancellations. In case $M_\phi$ is dominated by the
zero-temperature mass $m_\phi$, the resonance is therefore implied by the hierarchy between
$M_N$ and $m_\phi$, whereas otherwise, it is implied by the small Higgs-flavor violating
couplings $y$ and $\lambda$.

\section{Resonant Enhancement Required for the Observed Asymmetry}
\label{Sec:Obs:Bounds}

The observed value for the baryon asymmetry of the Universe is~\cite{Komatsu:2010fb}
\begin{align}
\label{BAU}
\eta_{\rm B}=\frac{n_{\rm B}}{n_\gamma}=(6.16\pm0.15)\times 10^{-10},
\end{align}
where $n_{\rm B}$ is the baryon number density and $n_\gamma$ the number of photons in
the cosmic microwave background. How small does $|M^2_{\phi11}-M^2_{\phi 22}|$ have to
be in order to account for the observed asymmetry for a given singlet neutrino mass
$M_N$? Leptogenesis for the flavor $i$
is typically most efficient in the regime between strong and
weak washout, and this transition occurs close to the point where
the product logarithm
$-W_{-1}$ takes its smallest real values~\cite{Buchmuller:2004nz},
{\it i.e.} when
\begin{align}
\notag
\frac25\left(\frac{2}{B_i}\right)^\frac25=\frac1{\rm e}
\Rightarrow
z_{{\rm f}i}=\frac52\,.
\end{align}
Moreover, the $CP$ asymmetry is maximized when $|Y_{\ell 1}|=|Y_{\ell 2}|$ and
the $CP$-violating phase is maximal.
The value $z_{{\rm f}i}=\frac52$ is then obtained for
\begin{align}
|Y_{i1}|^2=|Y_{i2}|^2=\frac{32\pi^4}{9}{\rm e}^{\frac52}\sqrt{2 g_\star/5}\frac{M_N}{m_{\rm Pl}}\,.
\end{align}
Up to the numerical factor, this relation is identical to the corresponding one
that determines the transiton between strong and weak washout regimes for
conventional Leptogenesis. Just as in the conventional scenario, it therefore
implies that the couplings $Y_{\ell i}$ are sufficiently small, such
that they do not jeopardize the upper bounds on the active neutrino masses,
even when the extra Higgs bosons acquire sizeable vacuum expectation values after
Electroweak Symmetry Breaking.
Besides, we assume that the couplings
of the lepton flavor $j$ are much larger than for the flavor $i$, {\it i.e.}
$|Y_{j1,2}|\gg|Y_{i1,2}|$,
such that the flavor $j$ suffers a stronger washout and its contribution to
the final asymmetry may be neglected. Under these assumptions, the first factor
in Eq.~(\ref{ass:fr}) is independent of the magnitude of $Y_{j1,2}$, because
$|Y_{j1}^* Y_{j2}|/{\rm tr}[YY^\dagger]\approx\frac 12$.

Putting these assumptions and approximations together, we evaluate Eq.~(\ref{ass:fr})
with the result
\begin{align}
\label{Yell:optimal}
Y_{\ell i}=9.8\times10^{-4}\frac{M_N^3}{\sqrt{g_\star}m_{\rm Pl}(M_{\phi 11}^2-M_{\phi 22}^2)}
=9.2\times10^{-5}\frac{M_N^3}{m_{\rm Pl}(M_{\phi 11}^2-M_{\phi 22}^2)}\,,
\end{align}
where we have taken $g_\star=114.75$ as the number of relativistic degrees of freedom
in the Standard Model with two additional Higgs doublets at high temperatures ($N_\phi=3$).

Now for definiteness, we assume that the Electroweak phase transition is of strongly first
order, such that~\cite{Harvey:1990qw,Chung:2008gv}
\begin{align}
\label{sphaleron:conversion}
\eta_{\rm B}=7.04 \times Y_{\ell i}\frac{24+4 N_\phi}{66+13 N_\phi}\,.
\end{align}
When the sphaleron freeze-out occurs after Electroweak Symmetry Breaking,
this conversion factor is slightly different, which should however not be relevant
within the accuracy of the present approximations.

From Eqs.~(\ref{BAU},\ref{Yell:optimal},\ref{sphaleron:conversion}),
we obtain the following upper bound on the splitting of the effective
({\it i.e.} including thermal corrections) Higgs masses
\begin{align}
|M_{\phi 11}^2-M_{\phi 22}^2|
<3.0\times 10^{-14} {\rm GeV}^2\frac{M_N^3}{(1\,{\rm GeV})^3}\,.
\end{align}
Under the assumption that the zero-temperature masses are small compared to
the effective masses, it is also illustrative to use Eq.~(\ref{M:decomp})
to turn this into a bound on the coupling constants
\begin{align}
\frac12
\left|
\sum\limits_k
(\lambda_{ki}^2-\lambda_{kj}^2)
\right|
+\frac{1}{12}
\left|
\sum\limits_l
(y_{li}^2-y_{lj}^2)
\right|
< 2.3\times 10^6\, \frac{M_N}{m_{\rm Pl}}\,.
\end{align}
It should be kept in mind that the scenario relies on flavor effects, {\it i.e.}
that there exist two effectively non-degenerate lepton flavors at the time of
Leptogenesis. Therefore, $M_N$ should be of order $10^{12}\,{\rm GeV}$ or smaller,
such that at least the $\tau$-Yukawa coupling is in equilibrium. (However, due to the
extended Higgs sector, it is also conceivable that the flavor-degeneracy may be already
broken at higher temperatures.)
It is interesting to
notice that when $M_N$ is of order $10^{12}\,{\rm GeV}$, no strong suppression of
the Higgs flavor-violating rates is required in order to explain the observed asymmetry.
We note that of course,
in the absence of a symmetry that suppresses flavor-changing neutral currents,
there are restrictive upper bounds on these Yukawa couplings or lower bounds
on the zero-temperature masses, {\it cf.} Ref.~\cite{Branco:2011iw} for a recent review.

The results of this Section illustrate that the resonant enhancement of $CP$-violation
in the present mechanism is of a quite different origin than in conventional scenarios
of Resonant Leptogenesis. While in the latter case, a small
mass difference of at least two singlet neutrinos must be imposed by virtue of
tuning or a symmetry, in the present scenario,
small zero-temperature Higgs masses compared to
the singlet neutrino mass and small Higgs flavor
violating couplings are required. The tuning responsible for
the smallness of the zero-temperature Higgs masses
can be identified with the tuning that is required in order to stabilize the Electroweak
scale compared to higher energy scales. The suppression of the Higgs-flavor
violating couplings of the
additional doublets is not well-motivated, unless one appeals to new symmetries,
but it is at the same time required by the suppression of
flavor changing neutral currents~\cite{Branco:2011iw}.

\section{Summary and Discussion}

Multiple Higgs doublets generically
contribute resonantly enhanced $CP$ violation to
flavored Leptogenesis. The resonance is typically controlled by the
the effective thermal mass degeneracy, that can be inferred from Eq.~(\ref{M:decomp}),
which is to be substituted into the expression for the final asymmetry~(\ref{ass:fr})
in the strong washout regime.

The intuitive picture of the present mechanism is that an out-of-equilibrium
distribution $\delta f_N$ induces off-diagonal correlations of the Higgs flavors.
Because gauge interactions are flavor blind, these do not directly damp
these correlations, but only force them into kinetic equilibrium.
Note that a partial cancellation of widths is also familiar from the decay asymmetry
in conventional Resonant Leptogenesis~\cite{Buchmuller:1997yu,Anisimov:2005hr,Garbrecht:2011aw,Garny:2011hg}.
The decay of the flavor-coherent
Higgs particles leads to a lepton flavor asymmetry.

While we calculate the asymmetry for $M_N\gg T$, we note that the present model
becomes similar to the one underlying Dirac Leptogenesis~\cite{Dick:1999je} when $M_N\to 0$. In
Ref.~\cite{Dick:1999je} however, the question of resonant enhancement is not addressed, while
Ref.~\cite{Bechinger:2009qk} misses that the
resonance is spoilt by large Yukawa couplings. In both,
Refs.~\cite{Dick:1999je,Bechinger:2009qk}, it is proposed that instead of $N$,
a right-handed electron
propagates in the loop, which would be in thermal equilibrium, such that no
asymmetry in $\ell$ can arise according to the discussion of Eq.~(\ref{coll:l}) and KMS.

The present work builds on two key aspects of more or less recently achieved
progress in Leptogenesis: the observation of the importance of
flavor~\cite{Endoh:2003mz,Pilaftsis:2005rv,Abada:2006fw,Nardi:2006fx} and the
reliable treatment of leading statistical
corrections~\cite{Buchmuller:2000nd,De Simone:2007rw,Garny:2009rv,Garny:2009qn,Anisimov:2010aq,Garny:2010nj,Beneke:2010wd,Beneke:2010dz,Garny:2010nz,Garbrecht:2010sz,Anisimov:2010dk}, that are crucial for the source terms~(\ref{S:Maxwell}) and~(\ref{S:explicit}).

While extra Higgs doublets may be experimentally
observable, the singlet neutrino generically is yet very heavy and too weakly coupled in order
to be discovered. It would therefore be interesting to conceive of models
that connect the present mechanism to the phenomenology of neutrino oscillations, in order
to obtain further constraints.
We close with two more speculative remarks: First, close-to degenerate states other
than Higgs doublets, that may be produced experimentally via gauge interactions, might
be connected with baryogenesis. Since their connection to the lepton sector must be either
indirect or mediated by non-renormalizable operators, the analysis of such models is presumably
less straightforward than in the present case. Second, it would be interesting to
re-consider Electroweak Baryogenesis from mixing particles, where it is often assumed
that the resonance is limited by gauge interactions, which is not the case for the present
scenario of Leptogenesis.

\subsection*{Acknowledgments}
The author is grateful to Martin Beneke and Marco Drewes for
valuable comments on the manuscript and acknowledges
support by the Gottfried Wilhelm Leibniz programme
of the Deutsche Forschungsgemeinschaft.

\end{document}